# A Machine Learning based Empirical Evaluation of Cyber Threat Actors High Level Attack Patterns over Low level Attack Patterns in Attributing Attacks

Umara Noor [a], Sawera Shahid [a], Rimsha Kanwal [a], Zahid Rashid [b]

[a] Department of Computer Science, Faculty of Computing and Information Technology
International Islamic University Islamabad (IIUI), Pakistan

[b] Technology Management Economics and Policy Program, College of Engineering, Seoul National University, 1 Gwanak-Ro, Gwanak-Gu, 08826, Seoul, South Korea

umara.zahid@iiu.edu.pk, sawera.mscs1109@iiu.edu.pk, rimsha.mscs1079@iiu.edu.pk, rashidzahid@snu.ac.kr

## ABSTRACT

*Cyber threat attribution is the process of identifying the actor of an attack incident in cyberspace. An accurate and timely threat attribution plays an important role in deterring future attacks by applying appropriate and timely defense mechanisms. Manual analysis of attack patterns gathered by honeypot deployments, intrusion detection systems, firewalls, and via trace-back procedures is still the preferred method of security analysts for cyber threat attribution. Such attack patterns are low-level Indicators of Compromise (IOC). They represent Tactics, Techniques, Procedures (TTP), and software tools used by the adversaries in their campaigns. The adversaries rarely re-use them. They can also be manipulated, resulting in false and unfair attribution. To empirically evaluate and compare the effectiveness of both kinds of IOC, there are two problems that need to be addressed. The first problem is that in recent research works, the ineffectiveness of low-level IOC for cyber threat attribution has been discussed intuitively. An empirical evaluation for the measure of the effectiveness of low-level IOC based on a real-world dataset is missing. The second problem is that the available dataset for high-level IOC has a single instance for each predictive class label that cannot be used directly for training machine learning models. To address these problems in this research work, we empirically evaluate the effectiveness of low-level IOC based on a real-world dataset that is specifically built for comparative analysis with high-level IOC. The experimental results show that the high-level IOC trained models effectively attribute cyberattacks with an accuracy of 95% as compared to the low-level IOC trained models where accuracy is 40%.*

***Keywords:** Cyber Threat Attribution, Threat Actor, Indicators of Compromise (IOC), Tactics Techniques and Procedures (TTP), Machine learning models, Cyber Threat Intelligence (CTI)*

## 1. INTRODUCTION

Cyber threat attribution is determining the identity or location of the adversary or the adversary's intermediary. It can be used in forensic inquiries by enterprises or in indictments by law enforcement agencies. This results in slowing down the pace of cyber-attacks by deterring future attempts, as people will be less inclined towards carrying out a cyber-crime if they know they will be caught. Governments and organizations require cyber threat attribution as a measure of proactive defense by identifying the adversary in the planning stage. For instance, the Cybersecurity Information Sharing Act (CISA) was enacted into law by the United States of America (USA) Congress in 2015 [1]. According to this rule, firms that have experienced cyber data leakage must provide information on threats to all parties, particularly their clients [2]. The attribution of cyber-attacks has changed throughout time in terms of perception and behavior. Almost a decade ago, attribution used to be focused on locating the perpetrator or middleman who launched a Distributed Denial of Service (DDOS) attack. This was designed for two purposes: first, to detect the culprit, and second, to prevent suspicious information from reaching its destination through IP trace-back. As a result, cyber-attack attribution worked as tracing-back and pinpointing the source of IP packets using routing tables. Trace-back procedures are discussed in depth in [3] [4]. The ability of IP address spoofing and anonymization constraint the effectiveness of source trace-back systems in identifying cyber adversaries. Reflection hosts, modest Time To Live (TTL) values, use of botnets, and increasing attack duration can all be used by a clever attacker to make identification difficult. With passing years, new forms of cyber-attacks known as Advanced Persistent Threats (APTs) have arisen, which include multi-stage multi-attack vector campaigns aimed at financial benefit, spying, and data theft against corporations, governments, and military. APTs are well-planned attacks that use advanced means to break into an infrastructure that is remotely administered via planting back doors. To timely detect and prevent an attack, the security community is paying attention towards recording and conveying detailed discussion of an event in Cyber Threat Intelligence (CTI) reports. Several standard languages have been proposed to represent CTI data. A brief introduction of the CTI tools and standards can be found in [5]. Among all such standards Structured Threat Information Expression (STIX) is the most comprehensive and widely adopted standard [6]. STIX stores the specifics of cyber-attacks as multi-level IOC characterized by observable, TTP, indicator, and exploit target components.

Based on the CTI, current practices of cyber threat detection mostly rely on identifying the perpetrator based on the IOC associated with that threat actor. There is a six-level hierarchy of IOC as discussed in [7]. The low-level indicators in the hierarchy represent file names, their hashes, IP addresses, and domain names. These indicators are simple and easy to detect and mitigate but they have a short life span that makes them less likely to be reused. Currently, low-level indicators are used for cyber threat attribution. The low-level IOC are transformed into firewall rules by the network administrator to prohibit fraudulent traffic. As discussed previously this attribution scheme based on low-level IOC has the same flaw as was in the attribution of a DDOS attack. The IP addresses can be easily spoofed or anonymized by the attacker which leads to inaccurate and biased attribution. Similarly, the malware hashes and domain names can be easily changed. Thus the problem of cyber threat attribution was redefined and evaluated for effectiveness according to the current demands of cyberinfrastructure and security information measurement and management architecture [8]. However, there are two main problems with the existing approach that need to be addressed. The first problem is that in recent research works [8], the ineffectiveness of low-level IOC for cyber threat attribution has been discussed intuitively. An empirical evaluation for the measure of the effectiveness of low-level IOC based on a real-world dataset is missing. The second problem is that the available dataset for high-level IOC has a single instance for each predictive class label that cannot be used directly for training machine learning models.

Based on the above research problems, the objective of this research work is to empirically evaluate and compare the effectiveness of both kinds of IOC based on real-world datasets. To address the first problem, we empirically evaluate the effectiveness of low-level IOC based on a real-world dataset that is specifically built for comparative analysis with high-level IOC. To address the second problem, an updated version of the high-level IOC dataset is built that can be used to train machine learning models and predict the culprit behind the cyber attack. Both datasets are provided to the research community for further research and exploration. The datasets are used to train different machine learning algorithms, such as, *(Random Forest (RF), Naïve Byes (NB), Naïve Byes kernel, Decision Tree (DT), K-nearest neighbors (KNN), Artificial Neural Network (ANN), Deep Learning, Gradient Boosted Trees, Generalized Linear Model, and Ensemble Learning Models).* The results show that high-level IOC-trained models effectively attribute cyber attacks (accuracy: 95%) as compared to low-level IOC (accuracy: 40%). Also, ANN is more effective in cyber attack attribution as compared to other algorithms.

The contributions of this research work are:

1. The low-level IOC training data set required to perform the experiment is built from publicly available CTI reports as there is no dataset available for low-level IOC with respect to attribution. The dataset associates cyber adversaries with their low-level IOC used in multiple instances.
2. A multi-instance high-level IOC training data set is built from publicly available adversaries' attack pattern taxonomy provided by MITRE [9]. The taxonomy is built from cyber-attack incidents described in security news articles, and technical reports by security service providers. They report the goals, motives, and capabilities of the attacker and sometimes associate them with the state sponsoring them.
3. Different statistical techniques are applied to both datasets to analyze their characteristics.
4. A comparative analysis of the effectiveness of machine learning models for both datasets is performed via empirical evaluation, which was intuitively done in previous literature.
5. The multi-instance high-level IOC dataset is used to attribute a recent cyber data breach incident on Red Cross for which the cyber threat actor is not known.

The paper is organized as follows. In the second section related work of the research domain in the context of cyber threat attribution and machine learning-based malware prediction are discussed. In the third section, the research methodology of cyber threat attribution is given. In the fourth section, the results of the experiment are given and discussed thoroughly. In the fifth section, a case study is presented that attributes a recent data breach incident based on their high-level IOC. In the sixth section, we highlight the factors limiting the scalability and feasibility of the proposed approach and suggest technological improvements to overcome them. Finally, in the seventh section, we conclude our proposed research work and provide a future plan.

## 2. RELATED WORK

Cyber threat attribution connected with high adversary assault patterns revealed in CTI documents is a relatively new study subject that cannot be readily compared and referenced to existing threat attribution methodologies. However, we review prior work linked to the challenge of attribution and the use of low-level IOCs for attack and malware detection in order to clarify the notion and interpretation of cyber threat attribution. In order to support the proposed approach, the significance of machine learning approaches is mentioned with respect to malware and intrusion detection systems.

To prevent DDOS attacks, source trace-back techniques are used which attribute cyber-attacks based on the IP addresses [3]. Identity spoofing, short TTLs, reflector hosts, and botnets as stepping stones are employed as features that make attribution challenging. Hunker et al. [4] suggest using information about an attack in order to attribute it. The organizations must share vulnerabilities, incident details, and new mechanisms for attribution. A set of actors (entities interested in attribution), the attributed objects, metrics to determine the degree of confidence in the attribution results and the entity providing attribution, an acceptable attribution policy, and the need to know privacy factors in attribution are among the attribution requirements highlighted [10]. Clark et al. [11] examine many types of attributions in the context of various attacks. The IP address is the primary observer of the situation. According to the conclusions of the study, IP trace-back is only effective in the event of a DOS/DDOS attack to stop the attack traffic deluge.

The OpenIOC framework reports low-level IOCs based on malware forensic analysis [22]. The research states that the low-level IOCs are firmly associated with the malware composition, i.e., hashes of binaries. They can be easily altered by polymorphism and metamorphism approaches composition. A detailed analysis of the cyber state criminals is reported by Kenneth et al [23]. According to the analysis, cybercriminals are globally classified into regions: Asia Pacific, Russia/ Eastern Europe, the Middle East, and the United States. It states that the ethnicity and cultures have a high impact on the behaviors and actions of cyber attackers. However, the analysis does not describe the attribution aspect experimentally. The role of machine learning in detecting intrusions and malware analysis is discussed in several research works. A comparative analysis of machine learning models to detect malware in the Android operating system is given in [24]. A blend of supervised and unsupervised machine-learning techniques for malware analysis is proposed in [25]. Artificial neural network is employed as an offline IDS to analyze cyber-attacks in Internet of Things (IoT) networks [26]. A deep neural network is detecting in malicious binaries by Saxe et al. [27], Saied et al. [28], and Kang et al. [29].

Noor et al. [8] proposed a new model to identify cyber threat actors' attack patterns automatically. They used Latent Semantic Index (LSI) search system to create their dataset of high-level IOCs from ATT&CK MITRE and corpus of intelligence reports. Data from intelligence reports covered events from (May2012 to FEB 2018) and 36 cyber threat actors' data is used. Also, they used machine learning algorithms to train the model. Accuracy of one model is 94%. However, LSI is not effective for cyber threat attribution and chances of false positive rate are high. Also, no comparison with low-level IOCs is provided using empirical evaluation and they claimed intuitively that low-level IOCs cannot identify cyber threat actors. Haddadpajo uh et al [31] proposed a model using fuzzy pattern tree, Multi-modal fuzzy, fuzzy C-mean partitioning to effectively address the cyber threat attribution. Basically, their research is about malware attribution. Accuracy of this proposed model is 95.2%. The dataset which is used in research consists of malware payloads of cyber threat actors. However, they considered only five cyber threat actors' data while 129 cyber threat actors' data is available on internet. Arun Warikoo [32] provides only conceptual idea for cyber threat attribution. The aim of this research is to help analyst within an organization for the effective attribution of cyber threat in the event of an attack. This conceptual triangle model is made up of three indicators: sector, tools and tactics, and techniques and procedures (TTPs).However, this model provides conceptual idea. Also, no empirical evaluation is performed on dataset. Naveen et al [33] proposed a model for cyber threat attribution using Word2Vec (SIMVER) and neural networks. The primary objective is to use the cyber threat intelligence reports with minor preprocessing, perform an effective attribution and to improve accuracy. The dataset used in the research was consisting of 12 APTs which is collected from various intelligence reports. Originally, the dataset is published in [34]. This proposed model achieves 86.5% accuracy for cyber threat attribution. However, only use data of 12 APTs (Small dataset). They have used domain specific word embedding technique but this is not explicitly extracting attack patterns from CTI documents. Themes using for attack pattern have not defined. Sentuna et al [35] proposed a model for cyber threat attribution using naïve byes posterior probability. The objective of this research is to enhance processing time and prediction accuracy of security against attack patterns of cyber threat actors. The accuracy of this proposed model is 95% and processing time is 0.021%. However, they only used 10 cyber threat actors' data which is small dataset. Also, noisy data in case of lost and poisoned attack patterns is not considered in the dataset.

Due to higher detection accuracy, real-time automated response and low computational resources, we also employ machine learning algorithms for solving the problem of accurate cyber-attack attribution.

## 3. Research Methodology

In this section, the methodology for the proposed research work is discussed. There are three steps: 1) Data Collection, and Dataset Preparation, 2) Statistical Analysis of Dataset, and 3) Cyber Threat Attribution. In the first section, the details of data collection for both kinds of IOCs are discussed. The sources from where the data is collected are mentioned. In the second section, the steps for dataset preparation are discussed. The features and predictive classes of both kinds of IOCs are elaborated using excerpts of the dataset. In the third section, the statistical analysis of the dataset is performed to reveal its intrinsic properties. In the fourth section, a brief review of the machine learning algorithms used for cyber threat attribution using both the datasets is given.

### 3.1 Data Collection and Dataset Preparation

To empirically evaluate the effectiveness of low-level IOC for cyber threat attribution, we couldn't find a compatible and ready to use dataset. Thus, we built the dataset of low-level IOC from CTI documents. The data for low-level IOC was collected from textual, publicly available and comprehensive CTI documents published by notable security solution providing IT and software enterprises such as Kaspersky [36, 37, 38, 39], Cylance [40, 41], Crowdstrike [42, 43], Novetta [44], Palo Alto Networks [45], Forcepoint [46], ThreatConnect [47], Cymmetria IT security company [48], F-secure IT security company [49], The citizen lab [50], Symantec Software Company [51], Fireeye cyber security company [52, 53], and IBM X-Force exchange [54].

A sample of a reconstructed CTI document about a cyber threat actor known as Deep Panda is shown in figure 1 [55]. The BlackBerry Cylance Threat Research Team are the authors of the document. The title of the document describes the cyber threat incident associated with deep panda. It can be seen that deep panda is also termed as shell crew by the security community. Along with the textual description, low-level IOCs are also present in the document in the form of file and malware hashes, IP addresses, and command and control server domains. Our goal is to collect these low-level IOCs from such textual CTI documents associated with a particular cyber threat actor that are used in their multiple campaigns.

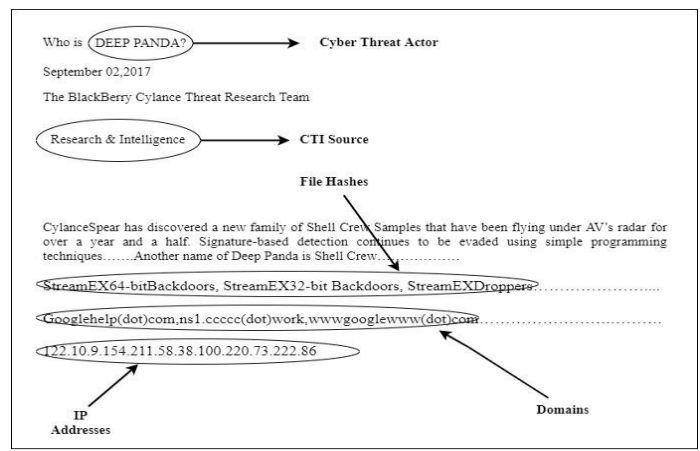

*Figure 1. Textual CTI document of "Deep Panda" Cyber Threat Actor*

Using this data, via empirical evaluation the effectiveness of low-level IOC in cyber threat attribution is determined. For this purpose, CTI documents related to 16 cyber threat actors were collected. A brief description of cyber threat actors is given in table 1. It can be seen that the cyber threat actors belong to different ethnicities. Their goals and motives distinguish them from one another, e.g., Carbanak has specifically targeted financial institutions and Poseidon is famous for black mailing the organizations. Also, the high-level IOCs of cyber threat actors are different.

Table 1. A brief detail of cyber threat actors

| Cyber Threat Actor | Description |
|---|---|
| Naikon [47] | Naikon is a suspected Chinese cyber threat group. It has been active in Southeast Asia since at least 2010. The main motive of this cyber threat group is to target government, military, and civic groups. |
| Deep Panda [42] | Deep Panda is a suspected Chinese threat group that has been linked to the government, defense, finance, and telecommunications sectors. Deep Panda is blamed for the hacking of Anthem, a healthcare organization. Shell Crew, WebMasters, KungFu Kittens, and PinkPanther are some of the other names. Based on attribution of both group names to the Anthem intrusion, Deep Panda also appears to be Black Vine. |
| DustStorm [41] | Dust Storm is a cyber-threat actor that has attacked a variety of businesses in Japan, South Korea, the United States, Europe, and Southeast Asia. |
| Suckfly [51] | Suckfly is a threat gang in China that has been active since at least 2014. |
| Carbanak [36, 38] | Carbanak has been linked to distinct organizations such as Cobalt Group and FIN7, both of which have employed Carbanak malware. Carbanak is a cybercriminal enterprise that has been targeting financial institutions with Carbanak malware since at least 2013. |
| Sandworm Team [49] | Sandworm Team is a destructive threat organization assigned to Russian military unit 74455 of the General Staff Main Intelligence Directorate (GRU) Main Center for Special Technologies (GTsST). Since at least 2009, this group has been active. |
| Lazarus Group [44] | The Lazarus Group is a North Korean state-sponsored cyber threat group. |
| Cleaver [40] | Cleaver is a threat group that has been linked to Iranian entities and is responsible for the Operation Cleaver activity. Cleaver appears to be tied to Threat Group 2889 based on strong circumstantial evidence (TG-2889). |
| Dark Hotel [37] | Dark hotel has been a suspected South Korean threat group that has targeted victims predominantly in East Asia. The name goal of this cyber threat group is to perform espionage operations and spear phishing campaigns by using peer-to-peer and file-sharing networks. |
| Poseidon Group [39] | The Poseidon Group is a Portuguese-speaking terrorist organization that has been active since 2005. The Poseidon Group has a history of blackmailing victims' firms into hiring the Poseidon Group as a security agency using information stolen from them. |
| APT30 [53] | APT30 is a cyber-threat outfit that has been linked to the Chinese government. While Naikon and APT30 have significant similarities, they do not appear to be identical duplicates. |
| Stealth Falcon [50] | Stealth Falcon is a threat group, circumstantial evidence shows a link between this organization and the government of the United Arab Emirates (UAE) since at least 2012.<br>The motive of this threat group is to target spyware assaults targeting Emirati journalists, activists, and dissidents. |
| GCMAN [38] | GCMAN is a threat group that primarily targets banks in order to transfer funds to e-currency providers. |
| APT 28 [43] | APT28 is a threat group linked to Russia's GRU military intelligence agency since at least 2004, this group has been active. APT28 allegedly hacked the Hillary Clinton campaign, the Democratic National Committee, and the Democratic Congressional Campaign Committee in 2016. |
| Patchwork [48] | Patchwork is a cyberespionage gang that was discovered for the first time in December 2015 circumstantial evidence implies it is pro-Indian or Indian. Patchwork has been seen focusing on industries that deal with diplomacy and government institutions. |

After collecting textual documents from the publicly available sources mentioned in the previous section, the low-level IOCs are manually extracted from them. The reason for manual extraction is the careful construction of the dataset without any discrepancies. In table 2, an excerpt of the low-level IOC dataset is shown. Each instance of the table represents a cyber threat actor and the associated malicious file hashes, IP addresses, and domains. The actual statistics of low-level IOCs for each cyber threat actor are given in section 3.2.

*Table 2. An excerpt of the low-level IOC dataset*

| Sr. No. | Cyber threat actor | File Hashes | IP Addresses | Domains |
|---|---|---|---|---|
| 1 | Deep Panda | • 2dce7fc3f52a692d8a84a0c182519133<br>• 1856a6a28621f241698e4e4287cba7c9<br>• de7500fc1065a081180841f32f06a537 | • 1.9.5.38,<br>• 202.86.190.3<br>• 202.86.190.3 | • sharepoint-vaeit.com<br>• gifas.asso.net |
| 2 | Naikon | • d085ba82824c1e61e93e113a705b8e9a<br>• 172fd9cce78de38d8cbcad605e3d6675 | • 202.86.190.3 | • cipta.kevins.pw linda.googlenow.in |
| 3 | Dust Storm | • 5961861d2b9f50d05055814e6bfd1c62<br>• 91b30719f8a4d02d4cf80c2e87753fa1 | | • adobeus.com<br>• moviestops.com |
| 4 | Sandworm | • bf9937489cb268f974d3527e877575b4 | • 195.16.88.6<br>• 93.115.27.57 | |
| 5 | Lazarus | • db4bbdc36a78a8807ad9b15a562515c4<br>• 2a04640352591b694b2d84be7b2b68f8 | | • www.htomega.com,mcm-yachtmanagement.com,www.junfac.com |
| 6 | Clever | • 01606d42c64e4d15ea07d4e1fbd0c40d<br>• 42714874F86FA9BD97E9BE460D7D<br>• 42E459D1D057BD937E0D00958E59 | • 78.109.194.96<br>• 217.11.17.96 | • DownloadsServers.com<br>• MicrosoftMiddleAst.com<br>• WindowsCentralUpdate.com |
| 7 | Dark Hotel | • 021685613fb739dec7303247212c3b09<br>• 1ee3dfce97ab318b416c1ba7463ee405 | | • 42world.net<br>• autoshop.hostmefree.org |
| 8 | Poseidon | • 2ce818518ca5fd03cbacb26173aa60ce<br>• f3499a9d9ce3de5dc10de3d7831d0938 | • 88.150.214.10 | • akamaihub.com<br>• mozillacdn.com |
| 9 | APT30 | • b4ae0004094b37a40978ef06f311a75e | | • www.bigfixtools.com |
| 10 | Fancy Bear | • bb909d9c27a509bf97cdc85268556ff5a | • 176.31.112.10 | • osce-military.org<br>• settings-yahoo.com |

High-level IOCs are extracted from ATT&CK Mitre [9]. It is a free repository of adversaries' tactics and techniques based on real-world real world security incidents. Tactics and techniques are termed as high-level IOC. They are taxonomized according to the post-compromise stages of the cyber kill chain model [56]. Cyber kill chain describes how an adversary or attacker launches an attack on a specific organization. Cyber kill chain has seven phases, i.e., reconnaissance, weaponization, delivery, exploitation, installation, command and control, and actions on objective. In the reconnaissance phase, an adversary finds a target and investigates vulnerabilities and weaknesses that can be exploited within the network. He may collect login credentials or other information, such as email addresses, user IDs, physical locations, software applications, and operating system characteristics. In the second phase the adversary designs an attack vector that can exploit a known weakness, such as remote access malware, ransomware, virus, or worm. The adversary may also set up back doors at this phase so that they can continue to access the system even if their original point of entry is recognized and closed by network administrators. The adversary starts the intrusion attempt in the delivery step. The particular actions they take will be determined by the type of attack they plan to launch. The attacker may send email attachments or a malicious link to encourage users to participate in the plan. The malicious code is executed within the victim's system during the exploitation phase. On the victim's computer, malware or another attack vector will be installed in the fifth phase. The attacker can use the virus to remotely control a device or identity within the target network under command and control phase. In the last phase, the attacker takes efforts to accomplish their objectives, which may involve destruction, encryption, and data theft.

The ATT&CK repository is constantly being updated via manual analysis of the publicly available CTI documents. Along with the high-level IOC, the repository also provides comprehensive details of the cyber threat actors examined by the security analysts globally. An interesting feature of ATT&CK is that it has exclusively identified the high-level IOC of the cyber threat actors which makes it a suitable source for cyber threat attribution. At the time of writing, ATT&CK has archived 567 techniques and 637 software tools related to 129 cyber threat actors. In [8] and [57], the older versions of ATT&CK are used. Each technique is represented as a text document with a unique identification that describes it along with the mechanisms for its detection and mitigation. The technique document also identifies the cyber threat actors who have used it in their campaigns.

Currently the resources provided by ATT&CK for the purpose of education and research do not have a ready to use high-level IOC dataset available. There are two challenges that need to be addressed. The first challenge is that the data about techniques and cyber threat actors is in separate web pages that are connected by hyperlinks. The machine readable ATT&CK data is accessible in the form of STIX [58]. The STIX version of ATT&CK, currently available on the website do not specify the cyber threat actors in the techniques documents available on github. This important connection between attack techniques and cyber threat actors is also missing in the excel spreadsheets provided by ATT&CK [59]. The first challenge is addressed by constructing the high-level IOC dataset manually from the ATT&CK website. An excerpt of the constructed high-level IOC dataset is given in table 3. The cyber threat actor is the class to be predicted. 'Drive by Compromise' and 'File and Directory Discovery' represents the high level IOC used by the cyber threat actor. Their IDs are 'T1189' and 'T1083'. It is mentioned earlier that high-level IOC are also termed as TTP. The 'China Chopper' and 'Bandook' represent the software tools with 'S0020' and 'S0234' IDs. The table entry of '1' shows that the cyber threat actor has used this technique in his campaigns and '0' shows that it is not used by the adversary.

*Table 3. An excerpt of high-level IOC dataset*

| Sr. No. | Cyber Threat Actor | Drive-by Compromise (T1189) | File and Directory Discovery (T1083) | …. | China Chopper (S0020) | Bandook (S0234) | …. |
|---|---|---|---|---|---|---|---|
| 1 | Dark Caracal | 1 | 1 | …. | 0 | 1 | …. |
| 2 | Dark hotel | 1 | 1 | …. | 0 | 1 | …. |
| 3 | Backdoor Diplomacy | 0 | 0 | …. | 1 | 0 | …. |
| 4 | Elder wood | 1 | 0 | …. | 0 | 0 | …. |
| 5 | Ferocious Kitten | 0 | 1 | …. | 1 | 1 | …. |
| 6 | Gall maker | 1 | 1 | …. | 0 | 1 | …. |
| 7 | Indigo Zebra | 0 | 0 | …. | 1 | 1 | …. |

The second challenge is that the constructed ATT&CK dataset is a single instance dataset. Each cyber threat actors' high level IOC are specified as a single instance or record in the dataset which cannot be used to train the machine learning models. To address this challenge, we synthesized the high-level IOC dataset to include multiple instances of a cyber threat actor. The justification behind the synthesis process is that a cyber threat actor does not use all the high-level IOC in a cyber threat incident. It can also be stated as it is not possible to detect all the high-level IOC of the cyber threat from a single incident. There is a possibility that some of the high-level IOC will be missed by the security analyst. Also, there is a possibility that the security analyst will detect some random high-level IOCs that are not part of the particular cyber threat and might belong to some other cyber threat. Such stray high-level IOCs can poison the actual detection process. These two possibilities are discussed in [57] as lost and poisoned TTP scenarios. Based on this fact, we have synthesized cyber threat actor's high-level IOC instances by adding noise to the original ATT&CK dataset. The noise represents the missed and poisoned high-level IOCs of the cyber threat actors. For each cyber threat actor, three more instances are added in the original dataset having 10%, 20%, and 30% noise. There were 129 cyber threat actors' instances in the original ATT&CK dataset. The synthesized dataset has 516 cyber threat actors' instances. The low-level and high-level IOC datasets used in this research work are provided for further use and analysis on github [60].

### 3.2 Statistical Analysis of Dataset

In this section, the statistical analysis of both low-level and high-level IOC datasets is discussed. The statistical analysis helps determine if the data distribution is normal or not. The reasons behind a specific state of a data distribution are mentioned. The statistics of low-level IOC are shown in table 4. The summary of the data distribution spread of file hashes, IP addresses, and malicious domains is shown in figure 2. It can be seen that the data of all three variables is not normally distributed. Mean and standard deviation are used for normally distributed data while median, and inter quartile range are used for non-normal distributions because they are not effected by outliers [61, 62].

*Table 4. Statisitcs of low-level IOC*

| Cyber Threat Actor | File Hashes | IP Addresses | Malicious Domains |
|---|---|---|---|
| Naikon | 22 | 0 | 52 |
| Deep Panda | 31 | 3 | 2 |
| Dust Storm | 54 | 0 | 61 |
| Suckfly | 18 | 1 | 7 |
| Carbanak | 37 | 81 | 35 |
| Sandworm | 73 | 13 | 0 |
| Lazarus | 76 | 0 | 4 |
| Cleaver | 41 | 11 | 15 |
| Monsoon | 40 | 5 | 39 |
| Dark hotel | 13 | 0 | 32 |
| Poseidon | 12 | 0 | 5 |
| APT30 | 12 | 0 | 8 |
| Stealth Falcon | 9 | 1 | 14 |
| GCMAN | 8 | 6 | 13 |
| Fancy Bear | 61 | 26 | 78 |
| Fin 6 | 5 | 0 | 0 |
| **Mean** | **32** | **9.1875** | **22.8125** |
| **Median** | **26.5** | **1** | **13.5** |
| **Interquartile Range** | **32.25** | **7.25** | **31.25** |

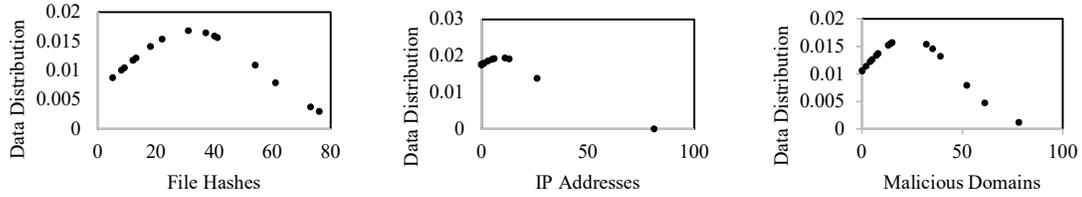

*Figure 2. Data Distribution of Low-Level IOC*

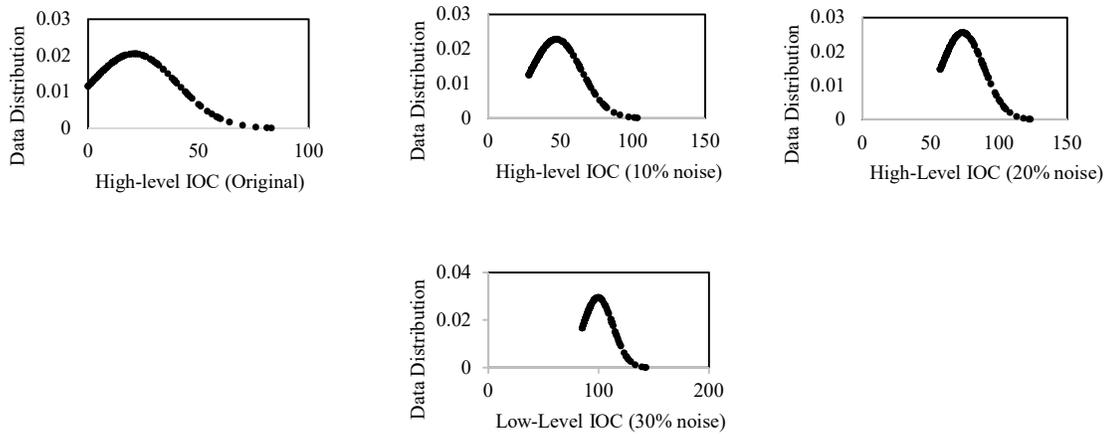

*Figure 3. Data Distribution of High-Level IOC for original and noisy datasets*

Therefore the median of file hashes, IP addresses, and malicious domain is 26.5, 1, and 13.5. The interquartile range of file hashes, IP addresses, and malicious domain is 32.25, 7.25, and 31.25.

The data distribution of the high level IOC dataset is shown in figure 3. It can be seen that the datasets are approximately normally distributed. Thus mean, and standard deviation are used as a measure of centrality and spread. The mean of high-level IOC for ATT&CK dataset, 10% noise, 20% noise, 30% noise is 20.71, 47.02, 73.28, and 99.38. The standard deviation value of high-level IOC for ATT&CK dataset, 10% noise, 20% noise, 30% noise is 19.45, 17.43, 15.56, and 13.49.

To estimate the difference between the original ATT&CK dataset and the synthesized dataset, Z test [63] is performed. Z test is used to determine if the mean of two datasets have significant different when the variances are known and the sample size is large. Here the sample size is 129 cyber threat actors. Here we have original ATT&CK data as the first group and the noisy data is the second group. The value of Z is-9.8551. The p-value is < .00001. It means that the result is significant at $p < .05$. This result shows that there is significant difference between the original ATT&CK dataset and synthesized dataset which is an evidence that the noise is randomly added to the original dataset and the synthesized dataset is unbiased.

## 3.3 Cyber Threat Attribution

In this phase, we train machine learning models using both low-level and high-level IOC datasets. The machine learning techniques used are: Naive Bayes [64], KNN [65], Decision Tree [66], Random Forest [67], Artificial Neural Network [68], Deep Learning [69], Generalized Linear Model [70], and Ensemble Learning (Voting, Stacking, Bagging, and Boosting) [71]. These models are used to anticipate the cyber threat actors behind an unseen cyber attack. In the following, a brief review of each technique is given.

### 3.3.1 Naive Bayes

Naive Bayes is a probabilistic classification system based on Thomas Bayes' posthumous theory of Bayesian Theorem [64]. It is employed in a broad range of classification problems. The Bayes theorem is a mathematical model that determines the conditional probability of an event occurring based on prior knowledge of factors that may be associated with the event. It has multiple variants such as Naive Bayes (Kernel). It is a class of algorithms that all follow the same principle: each pair of features to be categorized is independent of the others. The feature matrix and the response vector are the two elements of the dataset. All of the dataset's vectors (rows) are represented in the feature matrix, with each vector having the value of dependent features. The response vector holds the value of the class variable for each row of the feature matrix (prediction or output). It's incredibly beneficial when the inputs' dimensionality is high. It is simple and quick to forecast the test data set's class. It's also good at multi-class prediction. It doesn't take a lot of training data to discover intriguing insights, and when the data set is small, it may outperform complex machine learning models. Based on these features Naive Bayes classification algorithm intuitively seem suitable for both kinds of datasets under consideration in this research work.

### 3.3.2 KNN

The supervised machine learning algorithm k- Nearest Neighbor Classifier is beneficial for classification tasks [65]. Evelyn Fix and Joseph Hodges originated this concept in 1951 and Thomas Cover later expanded it. The KNN algorithm is based on feature similarity. It assigns a classification to a data point based on the classification of its neighbors. It keeps track of all available cases and categorizes new ones using a similarity matrix. The number of nearest neighbors to include in the majority voting process is denoted by K in KNN. The Elbow Approach is a commonly used method for obtaining the ideal value of k. With a positive integer k and a new sample, we select the k items in our data set that are closest to the new sample, and discover the most common classification of these entries. Then we assign this classification to the new sample. It is trivial to put into action. It can endure noisy training data. If the training data is large, it may be more effective.

### 3.3.3 Decision Tree and Random Forest

John Ross Quinlan developed a novel algorithm in 1986 known as decision trees [66]. They belong to the supervised learning category. They can be used to address problems involving regression and classification. Using the decision tree, we may depict any boolean expression on discrete characteristics. A decision tree is a graphical representation of all the possible outcomes of a decision based on a set of criteria. Each internal node in the decision tree algorithm represents a feature test, each leaf node represents a class label (a judgment made after computing all features), and branches represent feature combinations that lead to those class labels. The categorization rules are represented by the pathways from root to leaf. Using the decision tree, we may depict any boolean expression on discrete characteristics. The algorithm's focus is to develop a model that can predict the value of a target variable.

This approach is simple to comprehend, interpret, and visualize data processing. It requires minimum data preparation effort, and can handle both numerical and categorical data. It is further unaffected by non linear parameters. A limitation of decision trees is that they fail spectacularly on real data sets, leading them to overfit. Overfitting is a phenomenon in which a model performs well on training data but not on real or test data, resulting in a high variance. Decision trees have low bias, which means they fit well on your training data set, but high variance, which means they don't work well with data sets they haven't seen before. We still employ decision trees to solve this problem, but in a different form called random forest. Random forest is a technique that uses Ensemble Learning and is focused on the bagging algorithm [67]. It constructs as many trees as feasible on a sample of data and then integrates the results of all of the trees. As a corollary, the overfitting problem in decision trees is reduced, as is the variance, which enhances accuracy.

### 3.3.4 Artificial Neural Network and Deep Learning

In the late nineteenth and early twentieth millennia, the field of artificial neural networks (ANN) arose. An artificial neural network (ANN) is a biologically motivated computer model made up of many processing components that accept inputs and outputs based on their activation functions. ANN models replicate brain and nervous system electrical activity. A layer of input nodes and a layer of output nodes make up an ANN, which is connected by one or more layers of hidden nodes. By executing activation functions, input layer nodes send information to hidden layer nodes, while hidden layer nodes either activate or remain dormant depending on the evidence given. The evidence is weighted in the hidden layers, and when the value of a node or collection of nodes in the hidden layer hits a certain threshold, a value is sent to one or more nodes in the output layer [68]. The ability to build actual correlations between independent and dependent variables, as well as extract delicate information and complicated knowledge from representative data sets, is a distinguishing feature of ANN and it has the ability to handle the noisy data while A collection of statistical machine learning algorithms used to learn feature hierarchies, frequently based on artificial neural networks, is referred to as deep learning. The biological neuron serves as the motivation for neural networks, which are used to implement deep learning [69]. The human brain is made up of neurons, neural networks are built up of layers of nodes. Individual layer nodes are linked to nodes in neighboring layers. The number of layers in the network indicates that it is deeper. Deep learning's ability to collaborate with unstructured data is one of its most appealing features.

### 3.3.5 Generalized Linear Model

In 1972, John Nelder and Robert Wedderburn introduced the Generalized Linear Model, an advanced statistical modeling technique [70]. It's a catch-all phrase for a variety of models that allow the response variable to have an error distribution different from the normal distribution. The generalized linear model extends linear regression by allowing the linear model to be linked to the response variable via a link function and the size of each measurement's variance to be a function of its predicted value. The response variable does not need to be modified every time to have a normal distribution, unlike ordinary least square regression. This feature enhances its classification capability. Modeling allows for additional flexibility because selecting a link differs from selecting a random component.

### 3.3.6 Ensemble Learning

The ensemble learning approach is a collaborative decision-making process that combines the predictions of learnt classifiers to create new instances. Early study has demonstrated that ensemble classifiers are both empirically and logically more reliable than single part classifiers. To improve predictive efficiency, an ensemble model is a way for generating a predictive model by merging numerous models to tackle a single problem. It combines the output capability of more than one classifier to generate the final result which results in effective classification [71]. Voting, stacking, bagging and boosting (Adaboost) are some of the ensemble learning strategies employed.

In voting, multiple models produce predictions about a class in voting. The term "vote" refers to these forecasts. The majority of the models vote to determine the final prediction. The initial stage in this strategy is to generate several classification models from a training dataset. Each base model can be built by combining multiple splits of the same training dataset with the same algorithm, or by combining the same dataset with different algorithms, or by any other approach.

Stacking is the process of fitting multiple types of models to the same data and then using another model to learn how to integrate the predictions in the best way possible. On the test set, this model is utilized to make predictions.

Leo Breiman created the Bagging classifier in 1994 as an ensemble learning method for constructing a community of learners. Bagging is the process of fitting multiple decision trees to different samples of the same dataset and then averaging the results. Parallel ensemble is another name for bagging. During the learning phase, the base learners happen simultaneously.

Boosting includes sequentially adding ensemble members that correct prior model predictions and produces a weighted average of the predictions. Boosting strategies are made up of a collection of steps. From the original dataset, a subset is formed. All data points are given equal weighting at the start. On this subset, a baseline model is developed. On the entire dataset, this model is utilized to create predictions. The actual and forecasted numbers are combined to calculate the errors. Higher weights are assigned to data that were mistakenly anticipated. A new model is developed, and the dataset is used to make predictions. The weighted mean of all the models makes up the final model (weak learners). AdaBoost is implemented in this paper.

## 4. EXPERIMENTAL EVALUATION

To attribute cyber-attacks to their perpetrators, the machine learning models are trained with both the low-level and high-level IOC datasets. The experiments are performed on a PC with Intel Core m3-7Y30 processors and 1.00 GHz and 1.61 GHz processing speed, 8 GB of RAM running a 64 bit Windows 10. The machine learning models are evaluated using the k-fold cross validation resampling method. The method of cross validation approximate the strength of the machine learning model for unseen samples by splitting and grouping the data multiple times. The parameter 'k' determines the number of sub samples of the original data sample. The value of k is selected as 10. The whole dataset gets an equal chance in training and testing the machine learning model. The final results are obtained by averaging the results of the k-1 folds which is an unbiased and optimistic approach. The effectiveness of the machine learning models was evaluated with accuracy, precision, recall, and f-measure. The efficiency of the machine learning models was evaluated with the execution time of the experiment. The results of the effectiveness and efficiency of the machine learning models for the low-level IOC dataset is given in table 5. It can be seen that the highest accuracy achieved is 40.25% by ensemble learning model using stacking. In this model generalized linear model and another ensemble learning model, i.e., voting are stacked together. The reason is generalized linear model and voting approach has the second and third highest accuracies respectively as compared to other models. The accuracy, precision, recall, and f-measure for low-level IOC dataset is quiet low which clearly depicts that low-level IOC are incapable of effectively attributing cyber attacks. The execution time of low-level IOC trained models is reasonable.

*Table 5. Experimental Results for Effectiveness and Efficiency of low-level IOC*

| Machine Learning Algorithm | Accuracy | Precision | Recall | F-measure | Execution Time (hour: min : sec) |
|---|---|---|---|---|---|
| **Naïve Bayes** | 35.75% | 25.53% | 9.9% | 0.143 | 00:18 |
| **Naïve Bayes (Kernel)** | 35.75% | 25.53% | 9.76% | 0.141 | 00:05 |
| **K Nearest Neighbor** | 9.85% | 2.71% | 2.94% | 0.028 | 00:05 |
| **Decision Tree** | 19.52% | 0.53% | 2.7% | 0.009 | 00:05 |
| **Random Forest** | 26.6% | 3.89% | 5% | 0.043 | 00:08 |
| **Gradient Boosted Trees** | 33% | 10.93% | 10.17% | 0.105 | 03:43 |
| **Deep Learning** | 23.67% | 2.33% | 6.55% | 0.034 | 00:35 |
| **Generalized Linear Model** | 37.66% | 4.43% | 8.01% | 0.057 | 00:09 |
| **Ensemble Voting (Generalized Linear Model, Naïve Bayes, Random Forest)** | 35.75% | 8.06% | 7.85% | 0.079 | 00:31 |
| **Ensemble Stacking (Voting, Generalized Linear Model)** | **40.25%** | **5.76%** | **9.77%** | **0.072** | **00:46** |
| **Ensemble Stacking (Voting, Bagging)** | 28.84% | 5.34% | 5.89% | 0.056 | 01:02 |
| **Ensemble Bagging (Generalized linear Model)** | 37.66% | 4.61% | 8.03% | 0.058 | 01:38 |
| **Ensemble Ada boost (Generalized Linear Model)** | 8.12% | 0.22% | 2.7% | 0.004 | 00:10 |

The results of the effectiveness and efficiency of the machine learning models for the high-level IOC dataset is given in table 6. It can be seen that the highest accuracy achieved is 94.88% by Artificial Neural Network (ANN). The precision (93.95%), recall (94.88%), and f-measure (0.94) are also high as compared to other machine learning models. However, the efficiency evaluated by the model execution time is quiet low as compared to other models. It takes almost 6 hours to train the ANN model. However, once trained the model works efficiently for the unseen instances of dataset. The effectiveness of ANN is analyzed individually for each cyber threat actor. It is found that ANN is not able to attribute seven cyber threat actors, i.e., APT 16 [72], Bouncing golf [73], DragonOK [74], lotus blossom [45], NEODYMIUM [75], Scarlet Mimic [76], and Silver Terrier [77]. When the reason is examined, it is found that these cyber threat actors had minimum number of high-level IOC reported by the ATT&CK Mitre repository. APT 16 has one, Bouncing golf has two, DragonOK has two, lotus blossom has two, NEODYMIUM has one high-level IOC, Scarlet Mimic has five, and Silver Terrier has six high-level IOC reported in ATT&CK repository.

The second highest accuracy achieved is 89.44% by bagging approach of ensemble learning. In the bagging approach, the base machine learning model was random forest. The reason for selecting random forest is that it effectively attributed cyber attacks with an accuracy of 84.79% as compared to other approaches, i.e., naïve bayes (69.61%),

KNN (39.69%), and decision tree (20.78%). The execution time of bagging approach is 9 minutes approximately. The effectiveness of ensemble learning bagging approach is also analyzed individually for each cyber threat actor. It is found that it is not able to attribute eleven cyber threat actors, i.e., APT 16 [72], APT 30 [78], Black Oasis [79], Bouncing golf [73], DragonOK [74], Indigo Zebra [80], lotus blossom [45], NEODYMIUM [75], Orange worm [81], Scarlet Mimic [76], and Silver Terrier [77]. The first reason is the same reason that these cyber threat actors had minimum number of TTPs reported by the ATT&CK Mitre repository. The second reason is that the ensemble learning model is incapable of learning features of these cyber threat actors.

The accuracy, precision, recall, and f-measure for the synthesized high-level IOC dataset is promising which clearly depicts that high-level IOC are capable of effectively attributing cyber attacks even in case of lost and poisoned IOCs.

*Table 6. Experimental Results for Effectiveness and Evaluation of High-level IOC*

| Machine Learning Algorithm | Accuracy | Precision | Recall | F-measure | Execution Time (hour: min : sec) |
|---|---|---|---|---|---|
| Naïve Bayes | 69.61% | 83.23% | 69.61% | 0.76 | 0:01 |
| Naïve Bayes (Kernel) | 68.68% | 82.31% | 68.68% | 0.75 | 0:18 |
| K Nearest Neighbor | 39.69% | 61.37% | 39.69% | 0.48 | 0:01 |
| Decision Tree | 20.78% | 33.41% | 20.78% | 0.26 | 0:02 |
| Random Forest | 84.79% | 82.56% | 84.81% | 0.84 | 0:58 |
| Artificial Neural Network | **94.88%** | **93.95%** | **94.88%** | **0.94** | **5:59:42** |
| Deep Learning | 75.50% | 79.42% | 75.97% | 0.78 | 0:49 |
| Generalized Linear Model | 81.86% | 82.20% | 81.86% | 0.82 | 21:54 |
| Ensemble Voting (Random Forest, Generalized Linear Model) | 83.26% | 40.93% | 41.97% | 0.41 | 23:24 |
| Ensemble Stacking (Random Forest, Random Forest) | 86.5% | 84.86% | 86.51% | 0.86 | 1:57 |
| Ensemble Stacking (Bagging, Bagging) | 89.6% | 87.64% | 89.61% | 0.89 | 20:04 |
| Ensemble Bagging (Random Forest) | **89.44%** | **87.66%** | **89.46%** | **0.89** | **8:52** |
| Ensemble Ada boost (Random Forest) | 84.79% | 82.56% | 84.81% | 0.84 | 4:52 |

## 5. Red Cross Data Breach Case Study

The high-level IOC dataset is tested with a recent and unseen cyber threat actor's data breach incident on Red Cross [87]. In this data breach incident, the data of 0.5 million people was compromised. The data is about victims of conflict, war, and disasters. The data breach remained undetected for 70 days. However, the breached records are not publicly disclosed on any forum which depicts that the purpose of the cyber threat actor is espionage using stealth approaches. The sophistication of the attack shows that it is an Advanced Persistent Threat (APT). Currently, the cyber threat actor of this incident is unknown which makes this case a suitable study for analysis in this research work. In this section, we will attribute this data breach incident to its perpetrators using the high-level IOC of the incident. These attributes of the incident are provided as a test instance to the machine learning models trained using high-level IOC dataset discussed in section 4.

The high-level IOC of the Red Cross data breach incident are collected from publicly available international committee of Red Cross, news articles, and threat reports of security organizations [87--95]. There are 27 high-level IOC identified via manual analysis of the publicly available information about the incident. Here, a brief overview of high-level IOC is given. The cyber threat actors used sophisticated hacking tools which are employed in APT. They evaded security controls and remained stealth by using obfuscation techniques to hide their malware. The malware was customized according to the servers storing the data records. A critical vulnerability, CVE-2021-40539, in the authentication module lead to the network intrusion. Using this vulnerability, the cyber threat actors installed web shells in the server system and compromised administrator's credentials. Once getting the control of the server, hackers installed hidden tools on the server and accessed data records. Here, it can be seen that a major high-level IOC or tactic of the cyber threat actor is the stealth behavior. It is also known as "living off the land" approach [99]. In this tactic, the attackers make use of those tools that are legitimate and already installed on victim's machine. Another way is to run malicious codes directly in main memory. As a result fewer or even no new files are created in hard disk which can provide clues for detection. By using living off the land approaches it takes longer to detect data breaches.

In table 7, the cyber threat actor's prediction results of Red Cross data breach incident are shown. The machine learning models that have accuracy above 80% are considered for prediction. The ANN model which has highest accuracy of 95% predicted "Thrip" as cyber threat actor [96, 97]. The ensemble learning model using bagging and stacking approach which has the second and third highest accuracy of 89% and 86% respectively predicted "Threat Group 1314" as cyber threat actor [98]. Random forest, ensemble learning using voting, and Adaboost models predicted "FIN 10" as cyber threat actor [100]. The results depict that based on the high-level IOC information publicly available about the Red Cross data breach incident, the cyber threat actor predicted by the machine learning models may be one of these three actors, i.e., Thrip, Threat Group 1314, and FIN10 [96-100]. We closely examined the high-level IOC and previous data breach incidents of these cyber threat actors to find any possible link between these and the actor of Red Cross data breach. First, FIN 10 is examined. It is a financially motivated cyber threat actor whose incidents are related to financial fraud in casinos, and mining organizations. As mentioned earlier, the aim of the attackers of Red Cross data breach is espionage and there have been no report of any financial fraud and demand of ransom so far from the attackers so we belief that F10 is not the correct prediction. The prediction result may be due to the possibility of a few common high-level IOC in the related incidents.

Second, Threat Group 1314 is examined. The aims and motivations of the Threat Group 1314 are currently unknown. However, this group have used "living off the land" approaches to remain stealth which are also used by the cyber threat actor of Red Cross data breach incident. This important connection is identified by five of the machine learning models shown in table 7 which increases the possibility that the cyber threat actor of Red Cross data breach might be Threat Group 1314.

Finally, Thrip predicted by ANN is examined. As shown in section 4 that ANN has the highest accuracy of prediction. Thus it is expected that the prediction results of ANN model is more reliable than others. Thrip is an espionage group that has targeted communications of satellite, telecom industry, defense contractors, MapXtreme Geographic Information System (GIS) software, Google Earth Server, and Garmin imaging software. This shows an important connection with the cyber threat actor of Red Cross data breach incident whose aim is also espionage. The second important connection between the cyber threat actor of Red Cross data breach and Thrip is the employment of living off the land techniques to remain stealth and blend in with the victim's resources. Based on the publicly available information, both these important connections increases the possibility that the cyber threat actor of Red Cross data breach is none other than Thrip. Thus, in this case study the cyber threat actor predicted for Red Cross data breach incident is Thrip.

*Table 7. Cyber Threat Prediction for Red Cross Data Breach Incident*

| Machine Learning Algorithm | Cyber Threat Actor Prediction |
|---|---|
| Random Forest | FIN 10 |
| ANN | Thrip |
| Generalized Linear Model | Threat Group 1314 |
| Ensemble Bagging (Random Forest) | Threat Group 1314 |
| Ensemble Voting (Random Forest, Generalized Linear Model) | FIN 10 |
| Ensemble Stacking (Random Forest, Random Forest) | Threat Group 1314 |
| Ensemble Stacking (Random Forest, Generalized Linear Model) | Threat Group 1314 |
| Ensemble Stacking (Bagging (RF), Bagging (RF) | Threat Group 1314 |
| Ensemble Ada boost (Random Forest) | FIN10 |
| Ensemble Bagging (ANN) | Thrip |

## 6. LIMITATIONS AND SUGGESTIONS

The primary goal of this research is to elaborate on the importance of high-level IOC for cyber threat attribution .Our hypothesis was supported by the experiments with a good accuracy of the results. In order to actually implement the proposed solution in the real security environment there are certain limitations with the current CTI data and ATT&CK framework that need to be addressed. The first limitation is the difference in format of high-level IOC representation. High-level IOCs are normally represented as humanly understandable enumerations that are often long detailed sentences. While in ATT&CK repository high-level IOC are represented by a general concept with a unique ID assigned to each. This mapping between the high-level IOC requires human intervention. This limitation causes hurdle in deploying an automated solution for the proposed cyber threat attribution problem. Here we suggest to reference TTPs, software tools and threat actors in CTI documents with their ATT&CK assigned IDs just like the same way vulnerabilities are quoted and referenced with their specific CVE IDs [82] and low-level IOCs are addressed using common language framework i.e. OpenIOC [83] and Mitre CybOX [84]. The second limitation is there are certain TTPs that represent non-technical adversary traits e.g. the time at which adversary operates, language specific quotes and strings. In table 3, a few examples of such traits related to threat groups are given. Such traits play an important role in attributing cyber threats to their adversaries also some of them have high observability in the case of an attack as compared to other IOCs. Currently, ATT&CK framework lacks such traits. The third limitation is related to the completeness of CTI data. ATT&CK repository is the first effort to standardize the adversary's tactics techniques and common attack knowledge by categorizing TTPs and software tools to their associated adversaries. The problem is ATT&CK repository is not complete. It lacks important TTPs, tools and threat groups that can be found in other threat repositories such as IBM X-Force [54], Hail a Taxii [85] and Symantec [86]. Also, some important connections between these entities are also missing e.g. the famous Axiom threat group use HTTPs port 43 to bypass firewall and intrusion detection system. This TTP is not connected to the TTPs of Axiom group in the database. Similarly, the structured CTI data is manually managed and shared by different threat sources and platforms where certain intelligence is reported by a source and the other source lacks that. This limitation can be overcome with enriching CTI data with semantic data integration techniques. However, this is out of the scope of this research work.

*Table 2 Noteworthy and non-technical threat group attack patterns*

| Threat Group | Noteworthy Attack Patterns |
| --- | --- |
| Cleaver | Periodically rotates IP and Domain C&C addresses |
| Stealth Falcon | Use either fake (NGO or individual) identities or impersonate real ones for luring |
| Monsoon | Use phishing emails from the victim's mailing list provider |
| Patchwork | Lures are china related subjects and Pornography, initially determine the importance of the target then deploy full tool suite |
| Naikon | Geographical focus with country specific infrastructure, deploys proxies of the same country also use third country proxy C&C |
| Lazarus | Prefer IP over domain addresses for C&C |
| Fancy Bear | Working hours are Russian business hours, use Cyrillic keyboard |

## 7. CONCLUSION AND FUTURE WORK

In this research work to attribute cyber attacks to their perpetrators, the effectiveness of high-level attack patterns over low-level attack patterns is discussed. To empirically evaluate and compare the effectiveness of both kinds of IOC, the appropriate gold standard datasets are required. The dataset for low-level IOC for cyber threat actor attribution was not available. The standard high-level IOC dataset has a single instance for each predictive class label that cannot be used directly for training machine learning models. To address the first problem, the low-level IOC dataset is built from real-world cyber attack documents. To address the second problem, a synthesized version of the high-level IOC dataset is provided that can be used to train machine learning models and predict the culprit behind the cyber-attack. The datasets are empirically evaluated for their effectiveness and efficiency using machine learning models. Both the datasets are provided to the research community for further research and exploration. The experimental results show that the high-level IOC trained models effectively attribute cyberattacks with an accuracy of 95% as compared to the low-level IOC trained models where accuracy is 40%. Based on the results we conclude that high-level IOCs can attribute cyber threats to their perpetrators effectively. Currently the ATT&CK mitre repository is manually built that has a lot of missing high-level IOC for cyber threat actors. This effects the accuracy of attribution as well. In the future, we aim to automate the process of identifying high-level IOC of cyber threat actors from the textual cyber attack documents using clustering and non-parametric classification approaches to improve the effectiveness of the cyber threat attribution.

**Availability of data and materials**

The low-level and high-level cyber attack Indicators of Compromise (IOC) datasets are provided to the research community for further research and exploration on the following link. https://github.com/UmaraNoor/Cyber-Attack-Patterns-Dataset